\documentstyle[12pt,leqno]{article}

 \font\gotb eufm10 scaled \magstep1
 \newcommand{\bb}{\bibitem}
 \newcommand{\cc}{\cite}
 \newcommand{\lll}{\lambda}
 \newcommand{\lt}{\left}
 \newcommand{\rt}{\right}
 \newcommand{\vp}{\varphi}
\newcommand{\R}{\hat R}
 \newcommand{\Q}{\hat Q}
 \newcommand{\A}{\hat A}
\newcommand{\B}{\hat B}
 \newcommand{\AAA}{\hbox{\gotb A}}
 \newcommand{\beq}{\begin{equation}\label}
 \newcommand{\bea}{\begin{eqnarray}\label}
\newcommand{\eeq}{\end{equation}}
 \newcommand{\eea}{\end{eqnarray}}
\newcommand{\nn}{\\ \nonumber}
 \newcommand{\rr}[1]{(\ref{#1})}

  \author{D.A.Slavnov\thanks{Department of Physics, Moscow State University,
Moscow 119899, Russia. E-mail: slavnov@theory.npi.msu.su}} \title{Causality
 and probabilistic interpretation of quantum mechanics\thanks{Submitted to
 "Theoretical and Mathematical Physics"}} \date{\ } 
\begin{document}

\maketitle

\begin{abstract}

 It is shown that the probabilistic treatment of quantum mechanics can be
 coordinated with causality of all physical processes. The physical
 interpretation of quantum-mechanical phenomena such as process of
 measurement and collapse of quantum state is given.

\end{abstract}

 The purpose of the present paper is the substantiation of a thesis that the
probabilistic interpretation of quantum mechanics is quite compatible with
the supposition about unique causality of all physical processes. The
consideration is performed within the framework of the binary model (see
papers~\cc {slav}) of quantum mechanics in which it is supposed that there
are separate material carriers of corpuscular and wave properties in quantum
objects.

 Following the basic idea of the papers~\cc{slav} we shall consider that any
quantum object has dynamical and phase degrees of freedom.  Carriers of the
dynamical degrees of freedom are local. Further they will refer to as
nucleuses of the quantum object (to not confuse to atomic nucleuses).
Carriers of the phase degrees of freedom are fields, which further will refer
to as an information fields or a shell of a quantum object. These fields are
spread in the space.

 An elementary quantum object consists of one nucleus and a shell
(information field) which are coherent each other. It is possible to assume,
though it is not necessary, that nucleuses exist in pulsatory  (in the time
and the space) regime like some splashes of the information field. In this
case in microscopic scale a nucleus will not have continuous trajectory in
the space - time but in macroscopic scale the trajectory will exist.

 Action onto a quantum object can be dynamical and phase. The dynamical
action is accompanied by transmission of dynamical quantities. It acts onto
the dynamical degrees of freedom, i.e. onto the nucleuses. Respectively, the
nucleuses are responsible for corpuscular properties of the quantum object
and they contain the information about observables quantities in the latent
shape. The phase action is not accompanied by transmission of dynamical
quantities (or by very small transmission) and it acts on the phase degrees
of freedom. A classical measuring device reacts to the dynamical action of
the quantum object. Therefore the device reacts to an elementary quantum
object as onto one aggregate. But the concrete result of this reaction is
defined by structure of the shell of the quantum object. This structure
depends on state of the information field. Further we shall name it
physical state of the quantum object and we shall designate by symbol $ \vp$.

 Now we  try to formalize these physical ideas about the quantum object. In
order to take into account the latency of the information about observables
quantities we postulate, that the observables $ \B $ are Hermite elements of
noncommutative involute algebra (*-algebras) \AAA. Let's consider that to
each physical state $ \vp $ of a quantum object univalently there corresponds
a  functional on the algebra \AAA. This functional we shall designate by the
same symbol $ \vp $. The term "the physical state" will denote structure of
the information field, and the functional, corresponding to this structure.

 Thus, if $\B \in\AAA $ and $ \B^*=\B $ then $ \vp(\B)=B $ is a real
number (value of the observable $\B$) which is obtained in the {\it
concrete}  measurement.

 Let $ \{\Q \} $ is some maximal set of mutually commuting Hermite elements
of the algebra \AAA, i.e. $ \{\Q \} \subset\AAA $ and $$ \begin{array}{l}
\mbox { if } \Q_i, \; \Q_j \in \{\Q \} \mbox{ then } [\Q_i, \Q_j] =0; \\
\mbox{ if } \Q_i \in \{\Q \}, \quad \Q_j \in\AAA\mbox{ and } [\Q_i,  \Q_j] =0
\mbox{ then } \Q_j\in \{\Q \}; \\ \mbox{ if } \Q_i \in \{\Q  \}, \quad \Q_j
\notin \{\Q_j \} \mbox{ then } [\Q_i, \Q_j] \ne 0.   \end{array} $$

 The functional $ \vp $ maps the set $ \{\Q \} $ in a set of real numbers $$
\{\Q \} \stackrel{\vp}{\longrightarrow} \{Q = \vp (\Q) \}. $$ For the
different functionals $ \vp_i $, $ \vp_j $ the sets $ \{\vp_i (\Q) \} $, $
\{\vp_j (\Q) \} $ can differ and can coincide. If for all $ \Q \in \{\Q \} $
is valid $ \vp_i (\Q) = \vp_j (\Q) =Q $ then we shall term the physical
states $ \vp_i $ and $ \vp_j $ as $ \{Q \} $-equivalent. Let's denote by $
\{\vp \}_Q $ set of all $ \{Q \} $-equivalent physical states. Let's term
the set $ \{\vp \}_Q $ as a quantum state and we shall designate  $ \Psi_Q $.

 We shall take one fundamental supposition touching the physical state: each
physical state is unique, i.e. there are no two identical states in the world
and the physical states never repeat. It is possible to consider the physical
state is determined by all previous history of the concrete physical object
and this history for each object is individual. In particular, in two
different experiments we necessarily deal with two different physical states.
Different physical states $ \vp_i $, $ \vp_j $ correspond to different
functionals $ \vp_i (\;) $, $ \vp_j (\;) $, i.e. always there will be such
observable $ \B $, that $ \vp_i (\B) \ne\vp_j (\B) $.  From this supposition
follows, that each physical state $ \vp_i $ can be exhibited in form of the
functional $ \vp_i (\;) $ no more, than in one experiment.

 For each observable $ \A $ we shall introduce concept  "an actual state".
It is a physical state, in which the observable $ \A $ was  measured or will
be measured. A set of such states we shall denote by $ [\vp]^{\A} $. A set $
\{Q \} $ of equivalent states, actual for an observable $ \A $, we shall
designate $ \{\vp \}^{\A}_Q $. Following the standard quantum mechanics, we
shall consider that only mutually commuting observables can be measured in
one experiment.

 The functional $ \vp $ {\it is not linear}. However we shall require, that
on Hermite elements of algebra $\AAA$ a functional $\vp(\;)$, corresponding
to actual states, would satisfy to the following postulates: \bea {4} \quad
&1)&\vp(\lll \hat I)=\lll,\ \hat I\mbox{ is unity of algebra } \AAA,\ \lll
\mbox{ is real number}; \nn {} & 2)& \vp (\A +\B)=\vp(\A)+\vp(\B), \
\vp(\A\B)=\vp(\A)\vp(\B), \mbox{ if } [\A,\B]=0; \nn {} & 3)& \sup_\vp
\vp(\A^*\A)> 0, \mbox{ if} \A\ne 0; \qquad \vp (\A^*\A) =0, \mbox { if } \A=0;
 \nn{}& 4)& \mbox { for each set } \{Q \} \mbox { and each } \A\in\AAA \nn {}
&\quad&\mbox { there is } \lim_{n\to\infty} \frac {1} {n} \sum^n_{i=1} \vp_i
(\A) \equiv\Psi_Q(\A),\nn{}&\quad&\mbox { where } \{\vp_1, \dots, \vp_n \}
\mbox { is a random sample of the set} \{\vp\}_Q^{\A};\nn{}& 5)& \mbox { for
every } \A, \; \B\in\AAA \quad \Psi_Q (\A +\B) = \Psi (\A) + \Psi (\B). \eea

 We shall extend the functionals $ \vp (\;) $   onto anti-Hermitian elements
of the algebra $ \AAA $ with the help of the equality $ \vp (i\A) =i\vp (\A)
$. The  functional $ \Psi_Q (\;) $, appearing in the fourth postulate, has a
meaning of a functional $ \vp (\;) $ which is averaged over  all $ \{Q \}
$-equivalent actual states.  Symbolically it can be  represented in the form
of Monte-Carlo  integral   over the actual states: \beq{7}\Psi_Q (\A) =
\int_{\vp\in \{\vp \}_Q^{\A}} d\mu (\vp) \, \vp (\A). $$ We shall note that
$$\int_{\vp\in \{\vp \}_Q^{\A}} d\mu (\vp) =1. \eeq

 Let us  connect the functional $ \Psi_Q (\;) $ with each quantum state $
\Psi_Q $. The fourth postulate assumes that this functional does not depend
on a concrete random sample. Further the term "a quantum state $\Psi_Q$"
we shall use both for the set $ \{\vp \}_Q $ of physical states and for the
corresponding  functional $ \Psi_Q (\;) $ (the quantum average).

 Let $ \A^*\A =\B\in \{\Q \} $. If $ \vp\in \{\vp \}_Q $, then  $\vp
 (\A^*\A)=B$, where $B\in \{Q \} $. Therefore $$\lt.\Psi_Q (\A^*\A) =B =\vp
 (\A^*\A) \rt |_{\vp\in \{\vp \}_Q}. $$ From here $$ \| \A \|^2 \equiv\sup_Q
 \Psi_Q(\A^*\A)=\sup_{\vp\in[\vp]^{\A}}\vp(\A^*\A)>0,\mbox { if } \A\ne 0. $$
 As $ \Psi_Q (\A^*\B) $ is a linear (in $ \A^*\B $) positive semidefinite
functional then the Cauchy-Bunkyakovsky-Schwarz inequality is valid (see for
example \cc{blot})  $$ \Psi_Q (\A^*\B) \Psi_Q (\B^*\A) \le \Psi_Q (\A^*\A)
\Psi_Q  (\B^*\B).  $$ Therefore for $\Psi_Q  (\A^*\A) $ the postulates for
square of seminorm of the element $\A$ are fulfilled. Respectively it is
possible to accept $\|\A\|$ for the norm of $\A$. The algebra \AAA{}
becomes $C^* $ -algebra at  such definition of the norm . According to the
Gelfand-Naumark theorem (see \cc{blot}) every $C^*$-algebra can be realized
as algebra of linear operators in some Hilbert space. Thus, the proposed here
construction of quantum mechanics permit the standard realization.

 As against usual scheme of quantum mechanics in the proposed construction
one additional element is present. It is the physical state $ \vp $ and the
corresponding nonlinear functional $ \vp (\;) $. The functional $ \vp (\;) $
describes results of individual measurement in a concrete experiment, and the
functional  $ \Psi_Q(\;) $ describes average value of the observable in a
series of experiments, performed in identical conditions from the point of
view of the observer, i.e. at the same quantum state. The fact of existence
of the functionals $ \vp(\;) $ and $ \Psi_Q(\;) $, satisfying to the
enumerated postulates, is proved by all complex of quantum experiments.  The
 standard quantum mechanics is busy in problems describing by the functionals
 $ \Psi_Q (\;) $. However now single quantum phenomena take a great meaning.
 For example, such phenomena underlie an operation of quantum computer.
 Therefore it is desirable to supplement the formalism of the standard
 quantum mechanics by positions, which would allow considering the single
 quantum phenomena. On the other hand, it is extremely desirable, that such
 expanded formalism did not give rise to deductions, which would not agree
 with deductions of the standard quantum mechanics.

 Namely for sufficing this condition the postulate of uniqueness of each
physical state is accepted.  It follows this postulate  that the physical
state can not be univalently fixed. Really, to fix the functional $ \vp (\;)
$, we should know its value $ \vp (\B) $ for all independent observables $ \B
$. Physically it is not realizable. In one experiment we can find $ \vp
(\B_i) $ only for mutually commuting observables $ \B_i $, and in different
experiments we necessarily deal with different functionals $ \vp (\;) $.

The most hard fixing of the functional $ \vp (\;) $, which can physically be
realized, consists in reference it to some set $ \{\vp \}_Q $, i.e. to a
certain quantum state. For this purpose it is enough to perform measurements
of mutually commuting observables. It is possible to be restricted only by
independent measurements.  In principle it can be done in one experiment.
Thus we can not have the complete information about a physical state of a
concrete physical object in essence. The maximal observable and controllable
information on the physical object is concentrated in the quantum state.

 At the same time, the quantum states have some subjective element. From the
standard quantum mechanics it is well known that any quantum state can be
represented in form of superposition of quantum states which are fixed by one
maximal set $ \{\Q \} $ of mutually commuting observables. In the proposed
here construction it corresponds to the fact that one physical state can
belong to the different quantum states $ \{\vp \}_Q $ and $ \{\vp \}_R $.
I.e. $ \vp\in \{\vp \}_Q\cap\{\vp \}_Q $, where the states $ \{\vp \}_Q $
are classified by values of the set $ \{\Q \} $ of mutually commuting
observables $ \Q $, and $ \{\vp \}_R $ are classified by values of
observables $ \R\in \{\R \} $. The observables $ \Q $ and $ \R $ do not
commute among themselves. Then depending on what set ($ \{\Q \} $ or $ \{\R
\} $) we shall choose for classification the physical state $ \vp $ will be
referred either to the quantum state $ \{\vp \}_Q $ or to the quantum state
$\{\vp\}_R $.

 For example, let a spin-free particle decays into two particles $A $ and $B
$ with spins 1/2 which  scatter at large distance. Let's measure a projection
of spin onto the axis $z $ for the particle $A $. Let result will be $S_z (A)
$. Then using the conservation law, we can state that for the particle $B $
with absolute probability the projection of spin onto the axis $z $ is equal
$S_z (B) = -S_z (A) $. It denotes that the quantum state of the particle $B $
corresponds to such value of the projection of spin onto axis $z $. However
for the particle $A $ we could measure the projection of spin onto axis $x $.
Let result would be $S_x (A) $. Then we could state that the particle $B $ is
in the quantum state, which corresponds to the value $S_x (B) = -S_x (A) $ of
the observable $\hat S_x (B) $.

 As any physical operations with the particle $B $ are not fulfilled, the
physical state in both cases will be same
$\vp\in\{\vp\}_{-S_z(A)}\cap\{\vp\}_{-S_x(A)}$. The different quantum states
of the particle $B $ are related only to our subjective choice of the device
for measurement of the physical state of the particle $A $.

 This example is the description of the experiment proposed by Bohm~\cc{bom}
for demonstrating of the Einstein-Podolsky-Rosen paradox~\cc{epr}. In
proposed here treatment  any paradox is absent. The physical state $ \vp $
particles $B $, which is an objective reality, does not depend on our
manipulations with the remote particle $A $. Any transmission of an action on
distance is absent. The described experiment is an example of indirect
measurement at which the information about state of the quantum object is
obtained without physical action onto it. Usually interpretation of the
indirect experiment arouses the greatest difficulties, first of all, bound
with concept of collapse of quantum state (reduction of wave function). In
adduced example by our wish we "channelize" the particle $B $ into the
quantum state $ \{\vp \}_{-S_z (A)} $, or into the quantum state $ \{\vp
\}_{-S_x (A)} $, physically not acting onto the particle $B $. Such collapse
can be named  subjective (or passive) in contrast to objective (active)
collapse, which is related to the actual physical action onto the quantum
object. About the objective collapse we shall talk later. Here we shall note
that the subjective collapse is related to physical impossibility for us to
receive the complete objective information about the quantum object (the
complete information about the physical state). We should be content by the
partial information (quantum state).  It depends on our desire what concrete
part we prefer to be satisfied by.

It is possible also to use this experiment for demonstrating that, strictly
speaking, it would be possible to receive larger the  information about state
of quantum object, than ascertaining of its membership to this or that
quantum state. In the experiment we can measure the projection of spin onto
the axis $z $ for the particle $A $ and the projection onto the axis $x $ for
the particle $B $. In this case we can establish that the physical state $
\vp $ of the particles $B $ belongs to intersection of the corresponding
quantum states \beq {14} \vp\in\{\vp\}_{-S_z (A)}\cap\{\vp\}_{S_x (B)}. \eeq
But this information has specific character. It refer to the past, more
precisely, to the restricted interval in the past from the moment of decay of
the spin-free particle to the moment of measurement of the projection of spin
of the particle $B $. In this moment the information described by the
equation~\rr{14} will be garbled and will be useless for the further
monitoring (or control) of the particle $B $.

 Now we shall discuss dynamics and temporal evolution of quantum object. As
was already spoken, the action onto quantum object can be dynamical and
phase. By hypothesis elementary quantum object is nonlocal due to the shell.
The different parts of the quantum object can undergo different exterior
action and lose mutual coherence. In turn, it should result in disintegration
of the quantum object, since all constituent parts of the elementary quantum
object must be coherent by hypothesis. As the elementary quantum objects are
rather stable structures, there should be a cause, which hinders loss of the
coherence. Let's assume that such cause is a strong phase interaction within
the elementary quantum object. It recovers coherence. Let's assume also that
at the microscopic level the direct phase action onto quantum object is much
feebler than dynamical action.

 At the direct phase action the exterior objects act onto phase degrees of
freedom of quantum object directly. However indirect phase action is
possible. It is carried out as follows. The exterior objects dynamically act
onto dynamical degrees of freedom (onto nucleus) of the quantum object.
Further this action is transmitted to phase degrees of freedom through strong
interior phase interaction. It brings to  reorganization of the shell of the
quantum object, i.e. to modification of its physical state. Such phase action
is not weak as against dynamical.

 Neglecting the direct phase action, we shall consider that the evolution of
a physical state of quantum object is determined by dynamical action, and it
is controlled by a dynamical equation of motion, which is coordinated with
usual quantum-mechanical equation of motion. Namely we shall assume that the
physical state $ \vp $ evolves in the time so, that the functional,
corresponding to this state, $ \vp (\A) \quad (\A\in\AAA) $ varies as
follows: \beq{15} \vp_0 (\A) \to \vp_t (\A) \equiv\vp_0 (\A (t)), \eeq where
\beq{16} \frac{d\,\A(t)}{d\,t}=\frac{i}{\hbar}\Big[\hat H, \A (t) \Big],
\qquad \A (0) = \A. \eeq Here $ \hat H $ is a usual Hamiltonian (considered
as an element of the algebra \AAA) of the quantum object.

 The equations~\rr{15} and~\rr{16} quite  unequivocally describe temporal
evolution of the physical state. Therefore when  only  dynamical action is
accounted (it corresponds to that that  von Neumann~\cc{von} names as action
of the second  type) physical processes are strictly determinable. Other
matter, that with the help of observations we can determine the initial value
$ \vp_0 (\A) $  of the functional (physical state) only to within its
membership to some quantum state $ \{\vp \}_Q $. Therefore majority of our
predictions about the further observed dates for the considered quantum
object can be only probabilistic.

 Now we will turn to consideration of direct phase action. In the previous
reasoning we considered that they can be neglected. A situation however is
possible when it cannot be done. This situation is realized when the very
large number of exterior objects act the quantum object. As the nucleus is
local, it  feels action of small number of the exterior objects if the
long-range action is absent. As opposed to this, the shell having nonlocal
structure  feels action of the large number of the exterior objects. The mass
character of the action can cancel weakness of the separate action. It
happens only in that case when the separate weak actions do not cancel each
other. It is possible to assume that exactly such situation is realized at
action of a classical measuring device onto a quantum object. I.e.
distinctive feature of a measuring device is that its separate microscopic
elements  exerts synchronous direct phase action onto the quantum object.

 The typical classical measuring device comprises analyzer and detector. The
analyzer is a classical device with one inlet and several exits. In the
analyzer due to of direct phase action the united shell of quantum object
decomposes onto several coherent constituents. Symbolically we shall figure
it so: $$ \vp \to \vp_1\oplus\vp_2 \oplus\dots \oplus\vp_i \oplus\dots. $$
I.e. the united structure (the physical state $ \vp $)  decomposes onto the
direct (coherent) sum of constituents $ \vp_i $. Inside the analyzer everyone
evolves somehow , but all constituents preserve a mutual coherence.
Therefore, if in the further the constituents will incorporate they will be
able to interfere among themselves.

 The constituent $ \vp_i $ quits the analyzer through $i $-th exit. Everyone
($i $-th) exit corresponds to a particular value $A_i $ of certain observable
$ \A $ (or of several mutually commuting observables). Thus, the analyzer is
a point of branching of the initial physical state $ \vp $. The part of the
shell, falling into the $i $-th branch, belongs to a set $ [\vp]^{A_i} $,
which contains all information fields $ \vp ' $ such, that the corresponding
functionals satisfy to equality $ \vp'(\A) =A_i $.

 If only the dynamical action onto the quantum object is taken into account
point of branching of the shell is a point of bifurcation for motion of the
nucleus. In this point the interaction of the nucleus with the information
field plays a role of "random" force, which guides the nucleus along one of
the branch. Let's consider that the nucleus is retracted into the branch, for
which \beq {17} \vp \in [\vp]^{A_i}, \eeq where $ \vp $ is information field
of the quantum object $ before $ points of branching.

 Such motion through the analyzer is admissible for the shell not changing
structure. Let's consider that the nucleus should be in a resonance with
neighbouring part of the shell. Then such motion is admissible for nucleus
for which the resonant condition does not vary. Actually the structure of the
shell varies at the analyzer. Therefore equation~\rr {17} is necessary to
consider, as the requirement of an invariance of the resonant condition for
the nucleus at the point of bifurcation. The equation~\rr {17} is certain
condition of continuity for the motion of the nucleus. This equation
guarantees that at the point of bifurcation the evolution of the quantum
object is uniquely determinated by its physical state $ \vp $.

 The observable evolution has probabilistic character. Firstly, it is not
controlled by a dynamical equation of motion (the bifurcation point).
Secondly, the physical state $ \vp $ before the bifurcation point can be
fixed only to within  membership $ \vp $ to certain quantum state $ \{\vp
\}_Q $. Due to the equation~\rr {17}  the probability $W_i $ of falling  into
the $i$-th branch  for the nucleus  is determined by the equality \beq{18}
W_i = \int_{\vp\in \{\vp \}_Q\cap [\vp]^{A_i}} d\mu (\vp). \eeq

 Now we shall discuss the detector. It is a classical object, which has
strong dynamical and phase interaction with  quantum object. The detector is
in a macroscopically unstable state. As a result of dynamical action of
nucleus of the quantum object it goes out equilibrium. A catastrophic
process, which makes macroscopically observable result, develops in it.

 The phase action of quantum object onto the detector is proportioned onto
large number of microscopical constituents of the detector and does not give
 macroscopically observable effect. Thus, the detector macroscopically
reacts only to  nucleus of the quantum object, i.e. it reacts to the quantum
object as onto one aggregate.

 If the detector is located at the $i$-th branch  then it works with
probability $W_i $ (formula~\rr{18}). The nucleus of quantum object falls
into the $i$-th branch  with such probability. If the detector has worked, it
denotes $ \vp\in [\vp]^{A_i} $, i.e. $ \vp (\A) =A_i $. Thus, on the one
hand, value of the functional $ \vp (\;) $ really characterizes result of
individual measurement. On the other hand, using  formula~\rr{18} for average
value $ < A > $ of the observable $ \A $ we can receive $$ < A > = \sum_i
W_iA_i=\sum_i\int_{\vp\in\{\vp\}_Q\cap[\vp]^{A_i}} d\mu (\vp) \, \vp (\A) =
\int_{\vp\in \{\vp \}_Q} d\mu (\vp) \, \vp (\A) = \Psi_Q (\A). $$ It agrees
with deductions of the standard quantum mechanics and with  property~\rr{7}
of the functional $ \Psi_Q (\;) $.

 The inverse action of the detector onto the quantum object can go along two
scenarios. The first scenario is realized when the nucleus is at the branch
where there is the detector. The detector dynamically acts onto the nucleus
and strongly (directly and indirectly) onto that part of the shell of the
quantum object, which has fallen into the $i $-th branch. Due to this action
these parts of the quantum object lose coherence with those parts of the
shell, which have fallen into other branches. As a result, firstly, they lose
possibility  to interfere with the part of the shell, which has passed
through  the $i$-th branch.  Secondly, only this part of the shell remains in
the structure of the quantum object, as only it does not lose coherence with
the nucleus due to strong interior phase interaction.

Thus, there is sharp reorganization of the shell of the quantum object
(of its physical state). In standard quantum mechanics this phenomenon is
treated as a collapse of the quantum state. Here this phenomenon  can be
named objective or active collapse. By the exterior displays the objective
collapse is quite similar to the earlier described  passive collapse, but the
physical essences of these phenomena are completely different.

 It is necessary to note, that the modification of the physical state and, as
the consequence, its quantum state happens due to actual modification of the
part of the quantum object, which is at the detector. But not as a result of
vanishing (of reduction) of those parts, which are not at the detector.
Nothing happens with them. Nevertheless, they cease to be constituents of the
quantum object.
  In this case  the action of the detector onto the quantum object is
dynamical and phase. Either first type, or the second type of interaction can
predominate. If the overwhelming contribution gives the dynamical action then
this contribution can be described by the dynamical
equations~\rr{15},~\rr{16}.

 If overwhelming or the essential contribution
gives direct phase action then this contribution can not be described  by the
dynamical equations.  By terminology  of von Neumann it is interaction of the
first type. However  in this case (as opposed to the von Neumann's opinion )
the physical evolution of the quantum object is  uniquely determined by
structure of that part of the shell, which has hit the detector. Other
matter, that we have only  the information, which there is in the equation
$\vp_i\in \{\vp \}_Q\cap \{\vp \}^{A_i} $ for this part $ \vp_i $ of the
shell. Therefore we can do only probabilistic predictions. Most probably, if
the direct phase action plays main role  then  the modified part $ \vp'_i $
of the shell will belong to the set~$ \{\vp \}^{A_i} $, but it will cease to
belong to the set~$ \{\vp \}_Q $. In favour of  such supposition speaks
experiment, as  a particular quantum state is practically prepared  usually
so.

 Let's consider now second scenario, when the nucleus falls into that branch,
in which the detector is absent. For simplicity of reasoning we shall
consider that the analyzer has only two branches. The detector is located in
the second branch, and the nucleus falls into the first branch. In this case
the detector does not work (negative experiment). However the standard
quantum mechanics states that there is a collapse of a quantum state also.
Let's look, how it can be justified within the framework of considered here
model.

 In this case at the analyzer the field $ \vp $  decomposes onto coherent
constituents $ \vp_1 $ and $ \vp_2 $: $ \vp\to\vp_1\oplus\vp_2 $. The field
$ \vp_2 $  falls into operative zone of the detector. There this part of the
shell undergoes strong direct action of the detector. As a result of this
action $ \vp_2 $ loses a coherence with the nucleus and $ \vp_1 $. Therefore
$ \vp_2 $ ceases to be a part of the shell of the quantum object. Now the
quantum object will have  physical state $ \vp_1 $. There is  a modification
of the physical state of the quantum object. This modification is not
controlled by the dynamical equations (objective collapse of a quantum
state). In this case we quite definitely can state, that $ \vp_1\in
[\vp]^{A_1} $.

 Let's note that both at the first and second scenario,  on the one hand, as
a result of action of the detector there is a  (objective) collapse of the
quantum state of the quantum object, on the other hand, a long-range action
of the detector is absent.

 The proposed scheme of quantum mechanics gives  obvious and almost classical
explanation of the most fundamental quantum phenomena. The scheme is free
from any paradoxes. However there is a problem, maybe this scheme one of
variants of scheme with  hidden parameters. In a certain measure it is so,
but the reasonings, over which schemes with the hidden parameters are
rejected, is not correct in this case. The famous proof of von
Neumann~\cc{von} about impossibility of the hidden parameters in quantum
mechanics essentially founds on  linearity of quantum mechanics.  One of main
elements, functional $ \vp (\;) $, is not linear in the scheme, proposed
here. Therefore proof of von Neumann does not concern the present case.

 Other, not less famous, argument against schemes with the hidden parameters
is Bell inequality~\cc{bel}. We shall reproduce a typical deduction of this
inequality.  Let a quantum object $Q $ (particle with  spin 0 in the
elementary variant of experiment) decays into two objects $A $ and $B $
(particles with spins 1/2). The objects $A $ and $B $  scatter on large
distance and hit  detectors $D (A) $ and $D (B) $, respectively,  in which
the measurements are independent. The object $A $ has a set of observables $
\A_a $ (double projection of spin onto the direction $a $). The observables
corresponding to different values of an index $a $, are not simultaneously
measurable. Each of observables can take two values $ \pm1 $. In a concrete
experiment the device $D (A) $ measures an observable $\A_a $ with a
particular index $a $. For the object $B $ everything is similar.

 Let's assume that a quantum object $Q $ has a hidden parameter $ \lll $. In
each individual event the parameter $ \lll $ has a particular value. The
distribution of events according to the parameter $ \lll $ is characterized
by a measure $ \mu (\lll) $ with usual properties $ \mu (\lll) \ge0 $, $ \int
d\mu (\lll) =1 $.

  All magnitudes, connected with individual event, depend on the parameter $
\lll $. In particular, the values of observables $ \A_a $ and $ \B_b $,
obtained in a concrete experiment, are functions $A_a $, $B_b $ of the
parameter $ \lll $. For individual event the correlation of observables $
\A_a $ and $ \B_b $ is characterized by the magnitude $A_a (\lll) B_b (\lll)
$. The average value of the magnitude is referred to as correlation function
$E (a, b) $: $$ E (a, b) = \int d\mu (\lll) \, A_a (\lll) \, B_b (\lll). $$

  Giving various values to the indexes $a $ and $b $ and taking into account
that \beq{21} A_a (\lll) = \pm1, \quad B_b (\lll) = \pm1, \eeq we shall
obtain the following inequality \beq{22} |E (a, b) -E (a, b') | + |E (a', b)
+E (a', b') | \le \eeq $$ \le \int
d\mu(\lll)\,[|A_a(\lll)|\,|B_b(\lll)-B_{b'}(\lll)| + |A_{a'} (\lll) | \, |B_b
(\lll) +B_{b'} (\lll) |] = $$ $$ = \int d\mu (\lll) \, [ |B_b (\lll) -B_{b'}
(\lll) | + |B_b (\lll) +B_{b'} (\lll) |]. $$ In the right-hand side of the
formula \rr{22}, due to equalities \rr{21}, one of the expressions \beq{23} |
B_b (\lll) -B_{b'} (\lll) |, \qquad | B_b (\lll) + B_{b'} (\lll) | \eeq is
equal to zero, and another is equal to two for each value  of $ \lll $. From
here we obtain the Bell inequality \beq{24} |E (a, b) -E (a, b') | + |E(a',
b) +E (a', b') | \le 2.\eeq

 The correlation function $E (a, b) $ is easily calculated within the
framework of the standard quantum mechanics. In particular, when $A $ and $B
$ are particles with spin 1/2 \beq{25} E (a, b) = -\cos\theta_{ab}, \qquad
\theta_{ab} \mbox{ is an angle between } a \mbox{ and } b. \eeq It is easy
to verify that there are directions $a, b, a', b' $, for which
formulas~\rr{24} and~\rr{25} contradict each other.

 It would seem  it is possible to repeat this derivation in the proposed here
model, having made replacements of type $ A (\lll) \to \vp (\A) $, $B (\lll)
\to \vp (\B) $, $ \int d\mu (\lll) \dots \to \int_{\vp\in \{\vp \}_Q^{\A\B}}
d\mu (\vp) \vp (\dots) $. However this opinion is erroneous. For derivation
of the Bell inequality  it is essential, that in the left-hand side of the
inequality~\rr{22} it is possible to represent all terms  in the form of
united integral over one  parameter~$\lll $. It is not valid  for the
quantum average substituting this integral, as it is necessary to integrate
over actual states in it . The elements, appearing in different correlation
functions, $ \A_a\B_b $, $ \A_{a'} \B_b $, $ \A_a\B_{b'} $, $ \A_{a'} \B_{b'}
$, do not commute among themselves. Therefore sets of actual states,
corresponding to these operators, do not intersect.  In derivation of the
 inequality~\rr {24} we tacitly supposed that expression~\rr {23} exist for
 each $ \lll $. However there is no physical state $ \vp $, which would be by
 actual state both for the observable $\B_b$ and for the observable
 $\B_{b'}$.

In summary it is necessary especially to note, that the present paper is not at all attempt to formulate a new rival theory to quantum mechanics. The proposed scheme does not contradict any statement of the standard quantum mechanics. Maybe some theses gain slightly other physical interpretation. All deductions of the standard quantum mechanics are valid in the described scheme. At the same time there are additional elements (for example, functional $ \vp (\;) $, equality~\rr{4}) in this scheme. They allow to include individual events in the domain of its application.

Strictly speaking, the formalism of the standard quantum mechanics assumes that the classical relations are reproduced only for average values of quantum observables. Meantime the practice shows that such relations are reproduced in each individual experiment. Of course, it concerns only those observables, what can be measured in one experiment. In the proposed approach this fact is consequence of the equalities~\rr{4}.

 \end{document}